\documentclass[twocolumn,showpacs,preprintnumbers,amsmath,amssymb]{revtex4}

\usepackage{graphicx}
\usepackage{dcolumn}
\usepackage{bm}

\begin{document}

\preprint{APS/123-QED}

\title{Lack of self-averaging of the specific heat in the three-dimensional random-field Ising model}

\author{Anastasios Malakis}
\author{Nikolaos G. Fytas}
\affiliation{Department of Physics, Section of Solid State
Physics, University of Athens, Panepistimiopolis, GR 15784
Zografos, Athens, Greece}

\date{\today}

\begin{abstract}
We apply the recently developed critical minimum energy subspace
scheme for the investigation of the random-field Ising model. We
point out that this method is well suited for the study of this
model. The density of states is obtained via the Wang-Landau and
broad histogram methods in a unified implementation by employing
the N-fold version of the Wang-Landau scheme. The random-fields
are obtained from a bimodal distribution ($h_{i}=\pm2$), and the
scaling of the specific heat maxima is studied on cubic lattices
with sizes ranging from $L=4$ to $L=32$. Observing the finite-size
scaling behavior of the maxima of the specific heats we examine
the question of saturation of the specific heat. The lack of
self-averaging of this quantity is fully illustrated and it is
shown that this property may be related to the question mentioned
above.
\end{abstract}

 \maketitle

\section{Introduction}
\label{section1}

The random-field Ising model (RFIM)~\cite{imry75} is one of the
best studied glassy magnetic
models~\cite{belanger98,gofman96,newman96}, mainly because of its
interest as a simple frustrated system. In fact, it has been a
matter of conspicuous controversy over the last $15$ years, mainly
concerning the nature of its phase transitions. The RFIM
Hamiltonian is given by:
\begin{equation}
\label{eq:1}
\mathcal{H}=-J\sum_{<i,j>}S_{i}S_{j}-\sum_{i}h_{i}S_{i}
\end{equation}
where $S_{i}=\pm1$, $J>0$ is the nearest-neighbors ferromagnetic
interaction and the random-fields (RF's) are obtained from a
discrete distribution $h_{i}=\pm\Delta$, where $\Delta=2$ is the
disorder strength, also called randomness of the system.

The notion of dimensional reduction~\cite{parisi78} indicated that
the critical behavior of the RFIM in $d$ dimensions, at
sufficiently low randomness, should be identical to that of the
well-known normal Ising model in $d-2$ dimensions. On the other
hand, the droplet theory of domain wall energies in the
ferromagnetic state~\cite{grinstein82} suggested that a phase
transition should exist in three-dimensions (3D), for
finite-temperature and randomness. The puzzle has been cleared out
by Imbrie~\cite{imbrie84}, Schwartz~\cite{schwartz85} and Bricmont
and Kupiainen~\cite{bricmont87}. Their arguments strongly support
the view that a phase transition in 3D exists for sufficiently
small randomness $(\Delta_{c}\approx2.3)$~\cite{newman96}.

From the experimental point of view, a true realization of the
RFIM is hardly conceived. However, it has been shown that dilute
antiferromagnets in uniform external field (DAFF) represent
physical realizations of the RFIM~\cite{fishman79} and a number of
experiments investigated the phase transitions of such 3D
systems~\cite{belanger83}. These experiments have proven to be
very difficult and their interpretation doubtful due to the slow,
glassy dynamics of the system.

Although there exist several open questions about the phase
transition in the RFIM, it is now generally accepted that a new
fixed point controls the behavior of RF
ferromagnets~\cite{chayes86,cardy}. The significance of this for
the RFIM (in $d>2$) is that this new zero-temperature random fixed
point controls the whole critical line ($T_{c}(\Delta)$) and that
the RF's are always relevant. For disordered systems with weak
randomness which couples to the local energy (such as random-site
impurity or random-bond models) the crossover to a new random
fixed point, depends on the Harris
criterion~\cite{cardy,harris74}. According to this, the disorder
is relevant if the correlation length exponent of the pure model
($\nu=\nu_{pure}$) satisfies the condition $d\nu<2$ and this
condition may be stated, with the help of the hyperscaling
relation ($\alpha=2-d\nu$), as $\alpha>0$. Since the specific heat
exponent of the 3D Ising model is positive~\cite{malakis04}, weak
disorder should be expected to be relevant. In the case of the
RFIM the type of disorder is much more severe, since the
randomness couples to the local order parameter and the crossover
renormalization group eigenvalue is always positive~\cite{cardy}.
The inequality $\nu\geq 2/d$, derived by Chayes \emph{et
al.}~\cite{chayes86} for the correlation length exponent of a
generic disordered system ($\nu=\nu_{random}$) would imply, using
again hyperscaling, a negative specific heat exponent
($\alpha<0$). However, it is believed that hyperscaling is
violated in the RFIM and the specific heat exponent $\alpha$ is
related to $\nu$ by a modified hyperscaling law
$2-\alpha=(d-\theta)\nu$. The exponent $\theta$ characterizes the
scaling of the stiffness of the ordered phase at the critical
point~\cite{middleton02}. Thus, the specific heat exponent of the
RFIM is not restricted, by the above theoretical considerations,
to be negative~\cite{chayes86}.

The inconsistency of various estimations in the literature
concerning the critical exponent $\alpha$ is the origin of a long
lasting lively controversy, leaving open, so far, even the
question of divergence or saturation of the specific heat. The
specific heat of the RFIM can be experimentally measured and is of
considerable theoretical interest. Several Monte Carlo methods at
finite temperatures but also methods using ground state
configurations have been used to estimate the critical exponent
$\alpha$. Some of the ground state studies came up with strongly
negative values, ranging from $\alpha=-1.5$~\cite{rieger93} to
$\alpha=-0.5$~\cite{hartmann01,rieger95,nowak98}, whereas
Middleton and Fisher~\cite{middleton02} estimated in marked
disagreement $\alpha=-0.01\pm0.09$. Experiments on DAFF provided
evidence of a second order phase transition and a logarithmic
singularity for the specific heat~\cite{belanger85}. Recently,
Barber and Belanger~\cite{barber01} in their Monte Carlo study of
a DAFF model reported also that their specific heat curve closely
mimics a logarithmic peak. Moreover, it has been pointed out that
a strongly negative value of $\alpha$ causes serious difficulties
when it comes to finding a consistent set of scaling relations to
describe the critical behavior of the RFIM. These scaling
relations are consistent if one uses $\alpha\approx
0$~\cite{hartmann01}, which is also close to the experimental
value~\cite{belanger83}. Clearly more work is needed to understand
the specific heats behavior of the model. This important issue may
be intimately linked to the main physical finding of this paper:
the violation of self-averaging of the specific heat, illustrated
below in Sec.~\ref{section3}.

The RFIM has been studied numerically using
traditional~\cite{young86,ogielski86,rieger93,rieger95} but also
more sophisticated Monte Carlo techniques~\cite{newman96}.
However, the nature of the model demands enormous computer
resources. The equilibration of the system at low temperatures is
exponentially slow for large systems. Furthermore, in order to get
a good estimate of the mean properties of the system, it is
necessary to repeat the simulations for a large number of
realizations of RF's. In the present work we have applied the new
and popular Wang-Landau (WL)~\cite{wang01} and broad histogram
(BH)~\cite{oliveira96} methods to estimate the density of states
(DOS), $G(E)$, of the model. These methods have been employed in a
unified implementation using the Schulz~\textit{et al.} N-fold
version of the WL scheme~\cite{schulz01,malakis04b} and the energy
space was restricted using the recent critical minimum energy
subspace (CMES) technique~\cite{malakis04}.

The rest of the paper is organized as follows. In
Sec.~\ref{section2} we provide an outline of the numerical methods
which are involved in our calculations, including a brief
description of the WL and BH methods. The recently developed CMES
restriction is properly adapted and illustrated for the RFIM and
useful technical details are provided. In Sec.~\ref{section3} we
discuss the main conclusion of our work: the violation of
self-averaging of the specific heat of the RFIM by studying the
relevant probability distributions. The scaling behavior of the
pseudocritical temperatures and their sample-to-sample
fluctuations are also presented. Our conclusions are summarized in
Sec.~\ref{section4}.

\section{A numerical approach}
\label{section2}

We proceed to describe and appropriately adapt to the RFIM a
recently developed Monte Carlo approach, based on the WL algorithm
for estimating the DOS and using the idea of dominant energy
subspaces (CMES technique). Consider a particular RF realization.
Then, the specific heat and its peak are easily obtained with the
help of the usual statistical sums. The CMES
scheme~\cite{malakis04} uses only a small but dominant part
$(\widetilde{E}_{-},\widetilde{E}_{+})$ of the energy space
$(E_{min},E_{max})$ to determine the specific heat peaks. Let
$\widetilde{E}$ denote the value of energy producing the maximum
term in the partition function at the pseudocritical temperature
(corresponding to the specific heat peak) and $S(E)=\ln{G(E)}$ the
microcanonical entropy. Then, Eq.~(\ref{eq:2}) defines the CMES
approximation:
\begin{subequations}
\label{eq:2}
\begin{eqnarray}
\label{eq:2a}
C_{L}(\widetilde{E}_{-},\widetilde{E}_{+})=N^{-1}T^{-2}\left\{\widetilde{Z}^{-1}
\sum_{\widetilde{E}_{-}}^{\widetilde{E}_{+}}E^{2}\exp{[\widetilde{\Phi}(E)]}-\right.\nonumber\\\left.
\left(\widetilde{Z}^{-1}\sum_{\widetilde{E}_{-}}^{\widetilde{E}_{+}}E
\exp{[\widetilde{\Phi}(E)]}\right)^{2}\right\}
\end{eqnarray}
\begin{equation}
\label{eq:2b} \widetilde{\Phi}(E)=[S(E)-\beta
E]-\left[S(\widetilde{E})-\beta
\widetilde{E}\right],\;\;\widetilde{Z}=\sum_{\widetilde{E}_{-}}^{\widetilde{E}_{+}}\exp{[\widetilde{\Phi}(E)]}
\end{equation}
\end{subequations}
where $N=L^{3}$ and $(\widetilde{E}_{-},\widetilde{E}_{+})$ is the
minimum dominant subspace satisfying the following accuracy
criterion:
\begin{equation}
\label{eq:3}
\left|\frac{C_{L}(\widetilde{E}_{-},\widetilde{E}_{+})}{C_{L}(E_{min},E_{max})}-1\right|\leq\textit{r}
\end{equation}
with $r=10^{-6}$. This accuracy is extremely demanding compared to
the statistical errors produced by the DOS method and to the large
sample-to-sample fluctuations of the RFIM. An algorithmic approach
for specifying the CMES is described in Ref.~\cite{malakis04}.

Using an ensemble of $M (m=1,...,M)$ macroscopic samples of linear
size $L$ corresponding to different RF realizations we have
applied the described scheme in a broad energy space that covers
the overlap of the dominant energy subspaces for all RF's of the
ensemble. For a RF realization, say $m$, let us denote by
$(\widetilde{E}_{-,m},\widetilde{E}_{+,m})$ the location of the
dominant energy subspace defined by the above restriction, and by
$\Delta\widetilde{E}_{m}=\widetilde{E}_{+,m}-\widetilde{E}_{-,m}$
its extension. Our simulations were carried out in the broad
energy subspace, which covers at least the union of the individual
subspaces, i.e. $(E_{-[M]},E_{+[M]})\equiv
\bigcup_{m}(\widetilde{E}_{-,m},\widetilde{E}_{+,m})$, of total
extension $(\Delta E)_{[M]}=E_{+,[M]}-E_{-,[M]}$. Note that, the
total extension may be for large lattices several times larger
than the individual extensions. This practice has the advantage
that the approximation of the specific heat for a particular RF is
accurate in a wide temperature range, including its pseudocritical
temperature. Thus, the present implementation is not the most
efficient for the purposes of locating only the specific heat
peaks. However, this usage provides a more reliable alternative
for comparing the statistics of the specific heat peaks to the
averaged specific heat curve used in the literature (see for
example Ref.~\cite{rieger93} and also Sec.~\ref{section3}) and for
discussing the pathology of this quite common choice. Despite the
strong fluctuations of the energy value corresponding to the
maximum term of the partition function $Z$, the union of the CMES
for large samples of RF's is a relatively small subspace, compared
to that of the normal Ising model. Consider the case $L=16$. Then,
the energy levels used in Ref.~\cite{malakis04} for the normal
Ising model, counting from the ground state $(ie=1)$, are the
levels $(ie=1220-2410)$. Using an ensemble of $1000$ RF's the
union of the CMES was found to be the range of levels
$(ie=1-950)$, while our simulations were performed in a wider
range $(ie=1-1200)$. This is of the same order with that of the
normal Ising model and at least five times smaller than that of
the total energy space.
\begin{figure}[htbp]
\includegraphics*[width=8 cm]{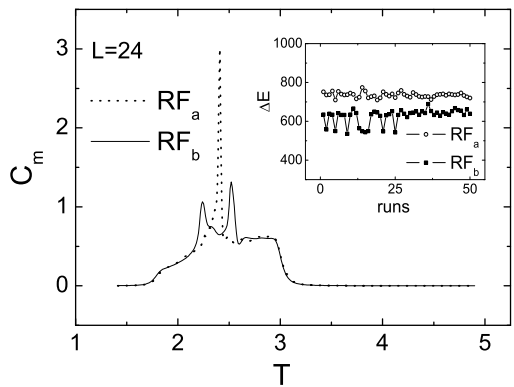}
\caption{\label{fig1}The specific heat for two characteristic
examples of RF's, $RF_{a}$ (dotted line) and $RF_{b}$ (solid
line). The specific heat curves were obtained using the average
DOS over the $50$ runs for each RF. The inset shows the
fluctuation of the extension of the individual dominant subspaces
over $50$ runs.}
\end{figure}

To conclude the above technical remarks, let us illustrate that
the efficiency of our method may be highly increased by studying
each RF realization in its own dominant subspace. The following
observations will be useful in subsequent studies of the RFIM or
analogous models. Fig.~\ref{fig1} shows results of an extensive WL
simulation of two particular RF's, labelled as $RF_{m=a}\equiv
RF_{a}$ and $RF_{m=b}\equiv RF_{b}$ in the figure, for a lattice
of linear size $L=24$. The simulation was repeated, for both RF's,
$50$ times in the energy subspace ($ie=200-2000$). The curves of
the specific heat shown were obtained from the average DOS over
the $50$ runs. Note however that, the union space used in our
simulations presented in Sec.~\ref{section3} (Fig.~\ref{fig5}) was
estimated over large ensembles of RF's. For instance, the union
space for $L=24$ and its extension was found to be of the order of
$2500$ energy levels ($ie=200-2700$), using an ensemble of $200$
RF's. Since there are RF's with one sharp peak and RF's with two
or more pronounced peaks we have chosen to show in Fig.~\ref{fig1}
two characteristic examples of RF's, $RF_{a}$ and $RF_{b}$. The
inset of Fig.~\ref{fig1} shows the fluctuation of the extension of
the individual dominant subspaces over the $50$ runs. There are
some points that one should observe from this figure. Firstly, in
both cases the extension of the CMES for a particular RF is much
smaller than the broad energy space used in the simulations.
Specifically, for the $RF_{a}$ the dominant energy subspace is
approximately of the order of $850$ energy levels ($ie=800-1650$),
which is almost three times smaller than the $2500$ levels of the
total union space. Secondly, the fluctuations of the extension
$\Delta\widetilde{E}_{b}$ of the $RF_{b}$ are more pronounced and
this is related to the existence of a secondary peak in the left
of the main peak. This secondary peak causes a stronger
fluctuation in the estimation of the end points of the
corresponding dominant subspaces. In any case, we could improve
the efficiency of our scheme by a factor of at least two (for the
case $L=24$), by carefully individualizing the used energy space
for simulating a particular RF. In fact, the fast early stages of
the WL process may be used as a prognostic method to approximately
locate the CMES of a particular RF, and this strategy may be an
indispensable ingredient in analogous future studies. Finally, an
entropic sampling study of the magnetic properties of the RFIM
using the CMES restrictive entropic scheme based on the
high-levels of the WL algorithm~\cite{malakis05} would be greatly
facilitated by such a strategy.

To determine the density of states, we have used the N-fold
version of the WL method as presented by Schulz~\textit{et
al.}~\cite{schulz01}, using $20$ iterations for the reduction
$(f_{j+1}=\sqrt{f_{j}},\;f_{1}=e)$ of the WL modification factor.
Our implementation is analogous to that presented in
Ref.~\cite{malakis04b}, where the first $13$ iterations follow the
simple WL scheme and the rest iterations $(j=14-20)$ use the
N-fold version of Schulz \textit{et al.}~\cite{schulz01}. We have
used a flatness criterion of $5\%$ for the energy
histogram~\cite{wang01,malakis04b}. For RF's $h_{i}=\pm2$ the
classes for the N-fold process are specified by the energy changes
$\Delta E_{n}=-16+4\cdot (n-1),\;n=1,2,...,9$. Using the part of
the simulation corresponding to the N-fold iterations $(j=14-20)$,
we have accumulated data corresponding to non-zero energy changes
in order to apply the well-known BH equation~\cite{oliveira96}:
$G(E)\langle N(E,E+\Delta E_{n})\rangle_{E}=G(E+\Delta
E_{n})\langle N(E+\Delta E_{n},E)\rangle_{E+\Delta E_{n}}$.
$N(E,E+\Delta E_{n})$ is the number of possible spin flip moves
from a microstate of energy $E$ to a macrostate with energy
$E+\Delta E_{n}$, which are known during the N-fold process. In
this way we have produced $4$ BH ($n=1,2,3,4$) approximations for
the DOS and the specific heat for each RF of the ensemble.

\section{Lack of self-averaging of the specific heat}
\label{section3}

For a disordered system we have to perform two distinct kinds of
averaging. For each sample, the usual thermal average has to be
carried out and then we have to average over the random
parameters. Let $C_{m}(T)$ denote the specific heat of a
particular realization $m$ in the ensemble of $M$ realizations of
RF's. The pseudocritical temperature $T_{L,m}^{\ast}$ will, of
course, depend on the realization of the RF. The location of the
corresponding peak is denoted by $(C_{m}^{\ast},T_{L,m}^{\ast})$
and the respective probability distributions by
$P_{L}(C_{m}^{\ast})$ and $P_{L}(T_{L,m}^{\ast})$.

\begin{figure}[htbp]
\includegraphics*[width=8 cm]{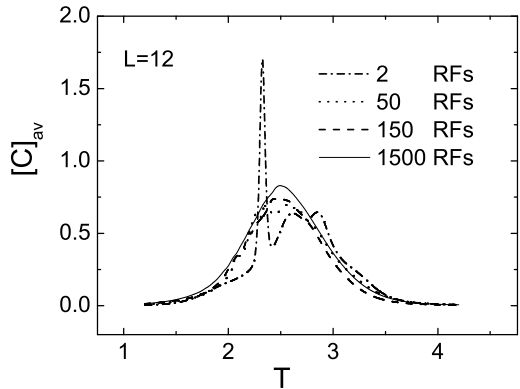}
\caption{\label{fig2}Averaged curves for various samples of RF's.}
\end{figure}
Rieger and Young~\cite{rieger93,rieger95} have studied the
following sample summation for the specific heat curves:
\begin{equation}
\label{eq:4} [C]_{av}=\frac{1}{M}\sum_{m=1}^{M}C_{m}(T)
\end{equation}
and the finite-size scaling behavior of the peak of this averaged
curve has been studied by assuming that the maximum
$[C]_{av}^{\ast}=max_{T}[C]_{av}$ and the corresponding
pseudocritical temperature $T_{L}^{\ast}$ obey the scaling laws:
\begin{subequations}
\label{eq:5}
\begin{equation}
\label{eq:5a} [C]_{av}^{\ast}\cong p+cL^{\alpha/\nu}
\end{equation}
\begin{equation}
\label{eq:5b} T_{L}^{\ast}\cong T_{c}+bL^{-1/\nu}
\end{equation}
\end{subequations}
Here we shall also examine the scaling of the sample averages of
the specific heat maxima and the pseudocritical temperatures,
defined by:
\begin{subequations}
\label{eq:6}
\begin{equation}
\label{eq:6a} [C_{m}^{\ast}]_{av}\equiv \frac{1}{M}\sum_{m}
C_{m}^{\ast}\cong
\widetilde{p}+\widetilde{c}L^{\widetilde{\alpha}/\widetilde{\nu}}
\end{equation}
\begin{equation}
\label{eq:6b} [T_{L,m}^{\ast}]_{av}\equiv \frac{1}{M}\sum_{m}
T_{L,m}^{\ast}\cong
\widetilde{T}_{c}+\widetilde{b}L^{-1/\widetilde{\nu}}
\end{equation}
\end{subequations}
The possibility of different exponents may be ultimately related
to the functional form of the distributions $P_{L}(C_{m}^{\ast})$
and $P_{L}(T_{L,m}^{\ast})$, whose behavior is decisive for the
comprehension of the critical behavior of the RFIM.

\begin{figure}[htbp]
\includegraphics*[width=8 cm]{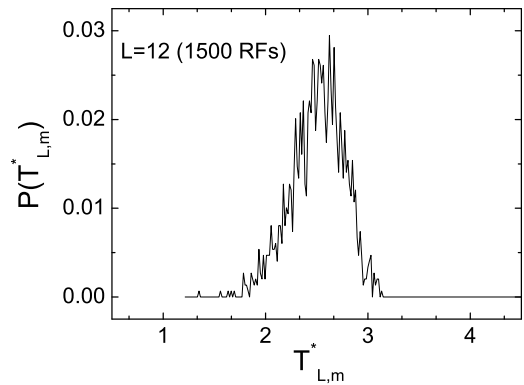}
\caption{\label{fig3}Fluctuation of pseudocritical temperature.}
\end{figure}

\begin{figure}[htbp]
\includegraphics*[width=8 cm]{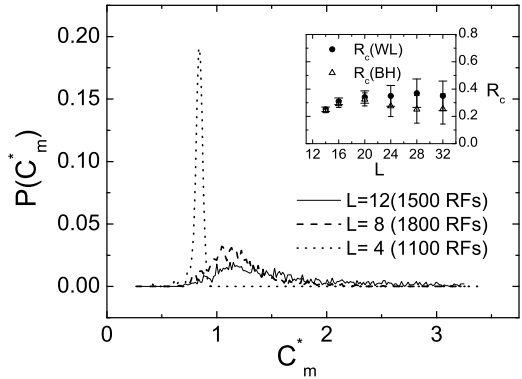}
\caption{\label{fig4}Broadening of probability distributions. The
inset presents finite-size evidence for the violation of
self-averaging. $R_{c}$ is defined and discussed in the text.}
\end{figure}

For a small number of RF's the averaged curve $[C]_{av}$ has
several local maxima reflecting a very strong sample-to-sample
fluctuation of the individual pseudocritical temperature.
Fig.~\ref{fig2} shows how the smoothness of this curve develops,
as we increase the number of RF's. Fig.~\ref{fig3} presents an
example of the probability distribution $P_{L}(T_{L,m}^{\ast})$.
Since the peaks are found in different locations, the averaging in
(\ref{eq:4}) wipes the particular peaks out. This explains why the
averaged curve does not represent the behavior of the most
probable, say $x$, realization of the RF's: $[C]_{av}(T)\neq
C_{x}(T)$. It also suggests the absence of self-averaging for the
specific heat of the present model, at least for the randomness
studied here. Fig.~\ref{fig4} illustrates the finite-size behavior
of the distribution $P_{L}(C_{m}^{\ast})$. Although for $L=4$ the
distribution is sharp, as $L$ increases the distribution broadens
so that there is a significant number of RF's having their peaks
higher, or lower, than the expected sample mean, defined in
Eq.~(\ref{eq:6a}). The above observations provide very strong
evidence that the real behavior of the RFIM is not appropriately
described by a possible misleading saturation of
$[C]_{av}^{\ast}$. The source of this problem is the severe
fluctuation of the pseudocritical temperatures and the lack of
self-averaging may be an important statement.

Broad distributions, with lack of self-averaging have been studied
also in other physical problems, such as in the well-known case of
the scaling theory of Anderson localization. There, it has been
shown that for a disorder electronic sample the conductance
distribution at the point of the metal-insulator transition (the
mobility edge) is so broad, that the conductance is not a
self-averaging quantity~\cite{cohen92}. Noteworthy that, the lack
of self-averaging appears to be a common property of disordered
systems at criticality and that besides the above mentioned
paradigm, one can find several examples of magnetic systems where
this feature is
present~\cite{honecker01,queiroz03,aharony96,wiseman98}. Actually,
when dealing with physical quantities that are characterized by
broad distributions, one must be mindful when attempting to define
a transition in terms of related averaged quantities. In this
sense, it seems that for the present model the common use of
$[C]_{av}^{\ast}$ may be a ``meaningless'' choice for a proper
description.
In order to discuss the significance of the above broad
probability distributions and to present a more convincing
finite-size scaling argument for the violation of self-averaging
in the thermodynamic limit, we have included as an inset in
Fig.~\ref{fig4} the ratio $R_{c}=V_{c}/[C_{m}^{\ast}]_{av}^{2}$,
where $V_{c}$ is the sample-to-sample variance of the average
(\ref{eq:6a}).
Both the WL and BH estimates are shown with their errors. The
variance $V_{c}$ was reduced, by eliminating the statistical (DOS)
errors, assumed to be of the order of the difference between the
two methods. The above defined normalized square width is a
measure characterizing the self-averaging property of a
system~\cite{aharony96,wiseman98,binder98}. This ratio appears to
tend to a constant $(R_{c}\rightarrow0.3)$, as can be seen from
the inset of Fig.~\ref{fig4}. Thus, according to the
literature~\cite{aharony96,wiseman98,binder98} the system is not
self-averaging and the corresponding distribution does not become
sharp in the thermodynamic limit.

\begin{figure}[htbp]
\includegraphics*[width=8 cm]{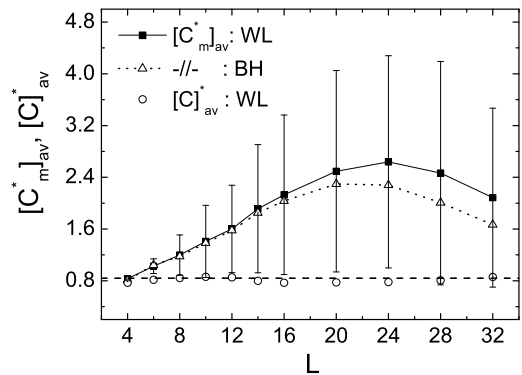}
\caption{\label{fig5}Size behavior of the averages
$[C_{m}^{\ast}]_{av}$ and $[C]_{av}^{\ast}$. The vertical bars
illustrate the order of the sample-to-sample fluctuations and
should not be confused with the small errors of the WL scheme.}
\end{figure}

The WL and BH estimates for $[C_{m}^{\ast}]_{av}$ appear in
Fig.~\ref{fig5}. For $L=4-20$ we have averaged over an ensemble of
$1000$ RF's and for $L>20$ over $200$ RF's. In this figure we show
the WL estimates and the mean (BH) of the $4$ BH estimates. The
same figure presents the size dependence of $[C]^{\ast}_{av}$.
Although for the range $L=4-20$ the behavior of the estimates is
convincing for their accuracy, an increase of statistical errors
is observed for larger sizes, depicted in the growing differences
between the WL and BH estimates. The estimates for $L=28$ and
$L=32$ appear to decline from the $L=4-20$ behavior and the
growing errors after $L=24$ make difficult a definite judgment for
the asymptotic behavior. Refinements of the WL scheme will be
favorable for these larger lattice sizes. This could be attempted
by using multiple measurements for each RF, by increasing the
final WL $j$-iteration and/or by introducing other refinements of
the WL algorithm, such as a separation $\mathcal{S}$ between
successive recordings~\cite{zhou05}. Our first attempt to increase
the WL $j$-iteration to $j=24$ for small samples of RF's indicated
that the level of $j=20$ leads to an underestimation of the sharp
peaks of the specific heat for most RF's. Nevertheless, this
observed underestimation was not of the order of the decline in
Fig.~\ref{fig5}, so it is possible that the model crossovers to
the conjectured saturation at these lattice sizes.

\begin{figure}[htbp]
\includegraphics*[width=8 cm]{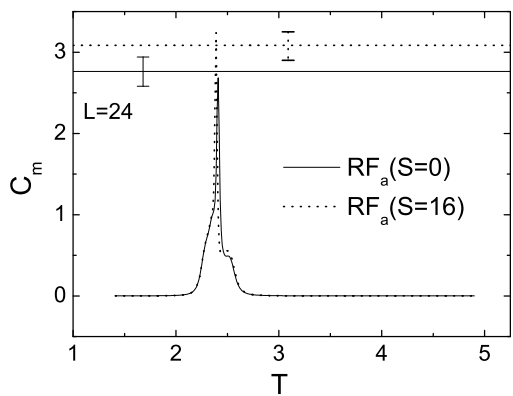}
\caption{\label{fig6}Illustration of the separation effect in the
specific heat of the $RF_{a}$. The specific heat curves shown were
calculated from the average DOS over the runs, while the
horizontal lines represent the mean values of the independent runs
for the specific heat peaks. The dotted line shows the case
$\mathcal{S}=16$, while the solid line the case $\mathcal{S}=0$.
The error bars illustrate the standard deviation (of the
independent peaks) over the $25$ (dotted) and $100$ (solid) WL
runs, corresponding to $\mathcal{S}=16$ and $\mathcal{S}=0$.}
\end{figure}

From our attempts to acquire a better comprehension of the reasons
for the above mentioned underestimation we also observed that this
aspect is quite strong for RF's with a sharp specific heat peak.
Note that such RF's are quite common and have been recently
discussed also by Wu and Machta~\cite{wu05}. This underestimation
may be observed also within the $j=20$ WL level by using multiple
measurements and also a separation $\mathcal{S}=16$ between
successive recordings of the accepted microstates of the WL
process. The separation refinement is generally believed to
improve the accuracy of the WL method~\cite{malakis05,zhou05}.
Fig.~\ref{fig6} illustrates its effect in a repeated application
using the $RF_{a}$, that appears also in Fig.~\ref{fig1}. From
Fig.~\ref{fig6} we observe that the effect of separation is to
increase the mean value of the maximum of the specific heat by an
amount which is of the same order with the standard deviation of
the statistical errors that one obtains by using multiple
measurements ($100$ independent WL runs) without separation.
The standard deviation of the new sample of multiple measurements
($25$ independent WL runs using separation) is also of the same
order, as shown in Fig.~\ref{fig6}. We note that, the WL sampling
in these multiple measurements was carried out in a energy
subspace which is slightly wider than the CMES of the $RF_{a}$
$(ie=700-1800)$ and not in the wider energy range used for the
simulation appearing in Fig.~\ref{fig1}. Comparing these two
figures (Fig.~\ref{fig1} and Fig.~\ref{fig6}) one can detect the
effects of different restrictions on the energy space. The
observed sadden decrease of the right tail of the specific heat (a
similar comment applies also for the left tail) in Fig.~\ref{fig1}
is an effect induced by the restriction imposed on the energy
space and appears in the neighborhood of $T\approx 3$. The further
restriction imposed in the new samplings (appearing in
Fig.~\ref{fig6}) is now reflected in the shift of the observed
sadden decrease in the neighborhood of $T\approx 2.5$. Before
attempting to simulate larger samples of RF's, other refinements
should be also tested, in order to obtain a more accurate and
efficient scheme. In any case, our study shows that there is a
large number of RF's with sharp peaks strongly fluctuating in
their pseudocritical temperatures and this generic property makes
the proposed CMES scheme the most appropriate alternative, despite
the accuracy and slowing down problems observed at the larger
sizes. Sharp peaks are usually missed by importance sampling, due
to an inadequate temperature scanning often used.

\begin{figure}[htbp]
\includegraphics*[width=8 cm]{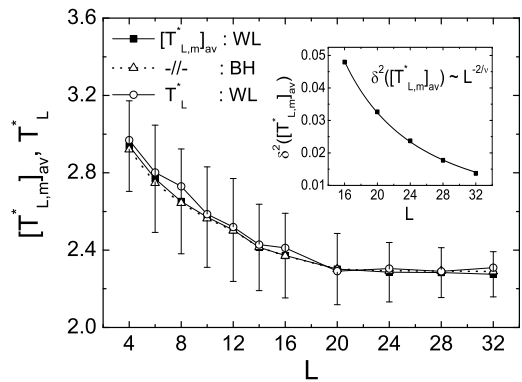}
\caption{\label{fig7}Size dependence of pseudocritical
temperatures. The vertical bars as in Fig.~\ref{fig5}. The inset
illustrates the scaling of the sample-to-sample variance of the
average $[T_{L,m}^{\ast}]_{av}$ (Eq.~(\ref{eq:6b})).}
\end{figure}

Fig.~\ref{fig7} illustrates that the two pseudocritical
temperatures, $[T_{L,m}^{\ast}]_{av}$ and $T_{L}^{\ast}$, tend to
the same limit ($\widetilde{T}_{c}=T_{c}$). The behavior of
$[T_{L,m}^{\ast}]_{av}$ is smoother than the behavior of
$T_{L}^{\ast}$, which is more sensitive to the sample size. Using
our data for $[T_{L,m}^{\ast}]_{av}$ we found a reasonably good
fit with $\widetilde{T}_{c}=2.03(18)$ and
$\widetilde{\nu}=1.31(18)$. This value is very close to the value
$\nu=1.37(9)$ found in Ref.~\cite{middleton02} and lies between
the values $1.0(1)$ of Ref.~\cite{nowak98} and the estimates
$1.6(3)$ and $1.4(2)$ of Ref.~\cite{rieger93}. The inset in
Fig.~\ref{fig7} illustrates the scaling of the sample-to-sample
variance of the average $[T_{L,m}^{\ast}]_{av}$ of
Eq.~(\ref{eq:6b}). Assuming that the square of these
sample-to-sample fluctuations scales with the linear size $L$
according to $\delta^{2}([T_{L,m}^{\ast}]_{av})\sim L^{-2/\nu}$,
we obtain, from the fit shown in the inset, the value
$\nu=1.18(15)$. This estimate is slightly smaller than the value
found above, which is in good agreement with the best estimate in
literature~\cite{middleton02}. The rather slow approach of the
fluctuations to zero is also an interesting finding. According to
Aharony and Harris~\cite{aharony96} and Wiseman and
Domany~\cite{wiseman98}, the fact that the square width of the
distribution of the sample dependent pseudocritical temperatures
scales with $L^{-2/\nu}$ and not with $L^{-d}$, when combined with
finite-size scaling~\cite{wiseman98}, is an indication of lack of
self-averaging of the random system. Therefore, our main
conclusion is reinforced and is also in conformity with the
results of Parisi and Sourlas~\cite{parisi02}. The numerical study
of these authors showed that the strong fluctuations of the 3D
RFIM produce a maximal violation of self-averaging for the
correlation length. It appears that the disorder present in the
RFIM brings about drastic effects and its strong non
self-averaging behavior includes also the specific heat, as
suggested in this paper.

It is quite possible that the above relevant aspect was overlooked
in previous finite-temperature studies but also in ground state
calculations. We think that, at least partly, this practice is
behind the existing controversial situation in the literature
concerning the behavior of the specific heat.  For instance, the
saturation of $[C]_{av}^{\ast}$ is evident from the very small
sizes and its behavior does not admit a finite-size scaling, but
rather appears as a random fluctuation around the value $0.84$, as
shown by the dashed line in Fig.~\ref{fig5}. The negative value
for the exponent $\alpha$ found, from the study of
$[C]_{av}^{\ast}$ in previous finite-temperature
studies~\cite{rieger93,rieger95} seems to us questionable. In
addition to all the reasons mentioned above, the very early and
clear saturation observed here and the possibility of a crossover
behavior of the model at larger lattice sizes are strong
indications that make us question the meaning of such a scaling
prediction. It appears that the behavior of $[C_{m}^{\ast}]_{av}$
incorporates more of the physical content of the model, although
its asymptotic behavior seems unsettled at the lattice sizes
studied here.
As pointed out earlier, the systematic errors of the WL scheme for
large lattice sizes are due to the practical ($j=20$) but not
fully converged usage of this algorithm in our simulations. This
option was dictated by the need of studying large samples of RF's.
We assume that the decline of the estimates observed here for
$L>24$ is stronger, as pointed out earlier, from the
underestimation observed by studying smaller samples of RF's and
using longer ($j=24$) runs. Then, it is quite obvious from
Fig.~\ref{fig5} that the true asymptotic behavior cannot be
observed at these lattice sizes, although its saturation seems to
be now plausible. In order to obtain a safe and sound estimation
of the large $L$-behavior, larger systems of at least of the order
of $L=60$ should be considered. This is an extremely demanding
computer project and will have to be postponed, until further
tests make available a more accurate and optimum refinement of the
presented scheme. Finally, let us point out that, our first
attempts to observe the behavior of the susceptibility of the 3D
RFIM via a recently proposed entropic scheme~\cite{malakis05},
suggested also an even stronger violation of self-averaging for
the magnetic properties of the system.

\section{Conclusions}
\label{section4}

In spite of many years of study, the conflicting situation in
literature concerning the divergence or saturation of the specific
heat of the RFIM is still an open important topic, necessary for a
better comprehension of the model. This problem was considered in
a completely new basis in this paper. The property of
self-averaging of the specific heat was addressed in a concise way
and its violation was explicitly shown by studying the relevant
probability distributions. This finding may lead to a better
theoretical and numerical approach of the problem. The scaling
behavior of the pseudocritical temperatures and their
sample-to-sample fluctuations were also presented, and found to
support a strong violation of the self-averaging property of the
system. The new ideas and numerical techniques utilized to tackle
the RFIM use as an essential ingredient the critical minimum
energy subspace scheme. We hope that the combination of algorithms
and techniques applied here will be useful in further numerical
studies of this and other similarly challenging problems.

\begin{acknowledgments}
This research was financially supported by EPEAEK/PYTHAGORAS under
Grant No. $70/3/7357$.
\end{acknowledgments}

{}

\end{document}